\documentclass{pasa}%

\usepackage[usenames,dvipsnames,svgnames,table]{xcolor}
\usepackage{float}
\usepackage{afterpage}

\title[]{How accurate are SuperCOSMOS positions?}
\author[Schaefer et al.]{Adam Schaefer$^1$, Richard Hunstead$^1$, 
\and Helen Johnston$^1$\\
\affil{$^1$Sydney Institute for Astronomy, School of Physics, University 
of Sydney, NSW 2006, Australia}}%
\jid{PASA}
\doi{10.1017/pas.\the\year.xxx}
\jyear{\the\year}

\begin{document}

\begin{abstract}

 Optical positions from the SuperCOSMOS Sky Survey have been compared in
detail with accurate radio positions that define the second realisation of
the International Celestial Reference Frame (ICRF2).  The comparison was
limited to the IIIaJ plates from the UK/AAO and Oschin (Palomar) Schmidt
telescopes.  A total of 1373 ICRF2 sources was used, with the sample
restricted to stellar objects brighter than $B_J=20$ and Galactic
latitudes $|b|>10^{\circ}$.  Position differences showed an rms scatter of
$0.16''$ in right ascension and declination.  While overall systematic
offsets were $<0.1''$ in each hemisphere, both the systematics and scatter
were greater in the north.

\end{abstract}
\maketitle%
\section{INTRODUCTION }
\label{sec:intro}

The accurate measurement of optical positions for celestial objects is
critical to all fields of astronomy, especially multiwavelength studies
involving radio, infrared, X-ray and $\gamma$-ray observations.
Traditionally, the celestial reference frame was linked to the Earth and
based on the positions of Galactic stars.  Stars, however, are not fixed
in space and the Earth is not a stable platform.  While the latter issue
can be overcome with satellite observations, stellar proper motions limit
the long-term accuracy of reference frames based on Galactic stars. The
solution has been to adopt a quasi-inertial reference frame based on the
radio positions of extragalactic objects, mostly quasars and BL Lac
objects, determined using very long-baseline interferometry (VLBI).

\subsection{Astrometric reference frames}

At present the most accurate astrometric measurements at optical
wavelengths are those made with the European Space Agency's (ESA)  
Hipparcos satellite \cite{perryman97}.  In 1995, the Hipparcos and Tycho
catalogues from this mission (mean epoch J1991.25) were tied to the
official IAU reference system known as the International Celestial
Reference System (ICRS). The ICRS is consistent with the FK5 (J2000)
system and its axes are defined by the VLBI positions of 212 ultra-compact
extragalactic sources constituting the defining sources \cite{ma98} for
the International Celestial Reference Frame (ICRF, now known as ICRF1);
these VLBI positions have errors $<$1~mas in each coordinate.  Since the
extragalactic objects were too faint for direct observation by Hipparcos,
the link between the two systems was made through additional ground-based
observations.  The estimated uncertainty in the link is dominated by
proper motion uncertainties of $\sim$0.25 mas/yr.

ICRF1 was based on a catalogue of 608 extragalactic sources which served
to define the ICRS from 1998 to 2009.  In 2009, ICRF1 was superseded by
the second realization of the International Celestial Reference Frame,
ICRF2 \cite{fey09}, which contains 295 defining sources, 97 of which were
defining sources in ICRF1.  Median positional accuracy for the 1217
sources (295 defining, 922 non-defining) is now $\sim$0.18 mas in each
coordinate.  ICRF2 has the advantage over ICRF1 of being distributed more
uniformly over the sky, although there is still a significant deficit of
sources in the south.  To be included in the ICRF2 defining list the
sources must (ideally) show no extended structure or structure variation
with time at the standard VLBI frequencies of 2.3 and 8.6 GHz.  However,
detailed monitoring of four ICRF2 sources with the Very Long Baseline Array (VLBA) over a year at 43, 23 and
8.6 GHz \cite{fom11} showed all of them to have multiple components.  
Although the relative positions of the identified radio cores were shown
to be stationary at a level of 0.02\ mas, two of the four ICRF2 sources
showed detectable motion in the direction of the radio jet.  Fomalont et
al.\ conclude that the influence of moving jet components sets a
fundamental limit on the present ICRF2 positions. In the optical, the
upcoming ESA astrometric space mission, {\it Gaia} \cite{perryman01,mk12},
will measure positions for $\sim$500,000 quasars to $G=20$ with precision
better than $\sim$0.2\ mas, allowing a direct link with ICRF2 at the
sub-mas scale.

\subsection{The SuperCOSMOS Sky Survey (SSS)}

The Schmidt telescope surveys of the northern and southern sky form an
invaluable resource for seeking optical counterparts of objects found at
other wavelengths, and for transferring a coordinate reference frame onto CCD images.  
They represent the last all-sky photographic surveys, and their digitized
versions continue to serve a wide range of astronomical projects.  The
SuperCOSMOS Sky Survey (hereafter SSS; Hambly et al. 2001c\footnote{http://www-wfau.roe.ac.uk/sss/} )was a high resolution (10\ $\mu$m
or 0.67 arcsec/pixel) digitization of the UK (southern) and Palomar
(northern)  Schmidt surveys using the SuperCOSMOS high-precision
microdensitometer. Hambly et al.\ \shortcite{ham98} have demonstrated
the capability of this machine to centroid well-exposed stellar images on
survey-grade plates to a precision of $\sim$0.5\ $\mu$m or 33 mas.

The surveys covered the sky in three colours --- blue (IIIaJ emulsion),
red (IIIaF emulsion), and near-infrared (IVN emulsion) --- with magnitudes
referred to as $B_J$, $R$ and $I$, respectively.  Limiting magnitudes are
$\sim$22.5 in $B_J$, $\sim$19 in $I$ and $\sim$21 in $R$.  Plates were 6.5$^\circ$ on a
side, and taken on 5$^\circ$ plate centres, giving substantial overlap
between adjacent plates to minimize edge effects.  As a result, objects 
near the plate boundaries were observed two or more times. In such cases 
the position adopted by the SSS \cite{ham01c} was taken to be the one closest to the plate centre,  
since position errors are known to increase with radial distance.

Positional accuracy of the SSS catalogue is claimed to be 0.1 arcsec for
objects with $B_J<19$.  The positions themselves are tied to the Hipparcos
reference frame via a second analysis of the Tycho catalogue 
using a more advanced reduction method\ \cite{hog00}.  The Tycho-2 catalogue, in turn,
is closely linked to ICRF1 \cite{ham01c}.

Given the complex distortions introduced by the bending of the glass
plates to the spherical focal surface of the Schmidt cameras, the goal of
our investigation was to determine the positional accuracy of SuperCOSMOS
over the whole sky through a detailed comparison with ICRF2.  Hambly et
al.\ \shortcite{ham01c} found that the smallest rms scatter between
SuperCOSMOS and ICRF1 for stellar objects was for the IIIaJ plates.  As a
result the comparisons with ICRF2 presented in this paper have been made
with SuperCOSMOS positions determined from the IIIaJ plates taken at the
1.2~m UK/AAO Schmidt telescope at Siding Spring Observatory in Australia
over the period 1974--1994, and the second epoch Palomar 
Observatory Sky Survey (POSS-II) plates from the 1.2~m Oschin
Schmidt telescope at Palomar Observatory in California, taken over the
period 1987--1999 \cite{ham01a}.  The UK Schmidt scans were made mostly on
the original survey plates, while the POSS-II scans were made on glass
atlas copies of the survey originals. The crossover declination between
the scans of the northern and southern surveys was at +2.5$^{\circ}$
(B1950).

Previous radio-optical position comparisons are presented in section 
\ref{prev_comp}, and the selection criteria for the present comparison are 
discussed in section \ref{source_sel}. Results are given in section 
\ref{results}, discussed in section \ref{dec_offsets} and summarised in 
section \ref{concl}.

\section{PREVIOUS COMPARISONS WITH ICRF}\label{prev_comp}

\subsection{USNO-A2.0}

Prior to the release of the SSS catalogue and images, the most accurate
digitization of the Schmidt plate surveys were those carried out at the US
Naval Observatory (USNO) by Monet et al.\ \shortcite{mon98}.  The
USNO-A2.0 catalogue of $>500\times 10^6$ objects, was based on scans of
POSS-I E and O plates north of declination $-18^{\circ}$ and UK Schmidt J
and European Southern Observatory (ESO) R plates south of $-18^{\circ}$.  
Digitization was at $0.9''$/pixel, and astrometric calibration was aligned
with ICRF through the USNO ACT Reference Catalog \cite{urb}.

A detailed comparison between USNO-A2.0 (and earlier digitized versions of
the Schmidt surveys) and ICRF1 was carried out by Deutsch
\shortcite{deu99}.  Of the 608 sources in ICRF1, 325 were rejected by
visual inspection as being too faint, too extended to obtain a reliable
centroid, or in crowded optical fields.  Mean offsets for the remaining
283 objects were $+0.04''$ in both RA and Dec, with maximum 
deviations of $\approx0.18''$ in each coordinate for the best 68\%.

\subsection{SuperCOSMOS}

Hambly et al.\ \shortcite{ham01c} used an approach similar to that used by
Deutsch \shortcite{deu99}, with a further constraint of limiting the
Galactic latitude to $|b|>30^{\circ}$.  This additional constraint reduced
the comparison sample to 110 stellar objects for the J/EJ survey and
somewhat fewer for the R and I surveys.  The most accurate results
came from the J/EJ surveys, with mean offsets of $-0.02''$ in both RA and
Dec for the 92 stellar objects with $B_J<20.0$ and maximum deviations of
$0.11''$ in each coordinate for the best 68\%.  It is important to 
note that the Hambly et al.\ \shortcite{ham01c} J/EJ comparison was 
limited to southern declinations.  Given the relatively small
numbers there was no attempt to explore the comparison as a function of
declination.

\section{SuperCOSMOS--ICRF2 COMPARISON}\label{source_sel}

Following the results obtained by Hambly et al.\ \shortcite{ham01c}, we
decided to restrict our comparison with ICRF2 to SSS positions
from the J/EJ surveys.  The radio positions used in the comparison come
from the 1217 sources in ICRF2 (295 defining, 922 non-defining), together
with a further 2197 ICRF2 compact sources comprising the VLBA Calibration
Source (VCS) list, all with positional accuracy $<1$\ mas \cite{fey09}.

Because of crowded star fields at low Galactic latitude, our comparison
was restricted to sources with $|b|>10^{\circ}$; Hambly et al.\
\shortcite{ham01c} adopted a more stringent limit of $|b|>30^{\circ}$.  A
further constraint was to limit the comparison to stellar objects brighter
than $B_J =20$, following Hambly et al.\ \shortcite{ham01c}.  Finally, to
exclude the majority of chance associations with foreground stars, matches
were sought only out to a radius of $1''$.  Despite this limit,
spectroscopic follow-up observations \cite{tit11,tit13} identified several
cases where the optical field of an ICRF2 source was obscured by a
closely-aligned foreground star;  we believe, however, that such cases are
rare and unlikely to affect the statistical results from this study.

Of the 3414 sources comprising the ICRF2, 853 were either not detected in
the SSS J/EJ catalogue or 
were unlikely identifications with position offsets $>1''$.  A further 178
sources with $|b|<10^{\circ}$ were rejected, along with 378 objects
catalogued as extended and likely to be galaxies (or blends), with
correspondingly larger centroiding errors.  The faint magnitude limit of
$B_J=20$, together with a bright limit of $B_J=15$, where image saturation
can affect the positional accuracy, eliminated a further 632 sources.  
The final sample contained 1373 ICRF2 matches, of which 187 were from the
set of 295 defining sources.  The much larger size of our sample in 
comparison
with previous studies by Deutsch \shortcite{deu99} and Hambly et al.\
\shortcite{ham01c} has allowed us to examine positional offsets as a
function of declination.

\begin{figure}[h]
 \begin{center} 
 \includegraphics[width=\columnwidth,clip=]{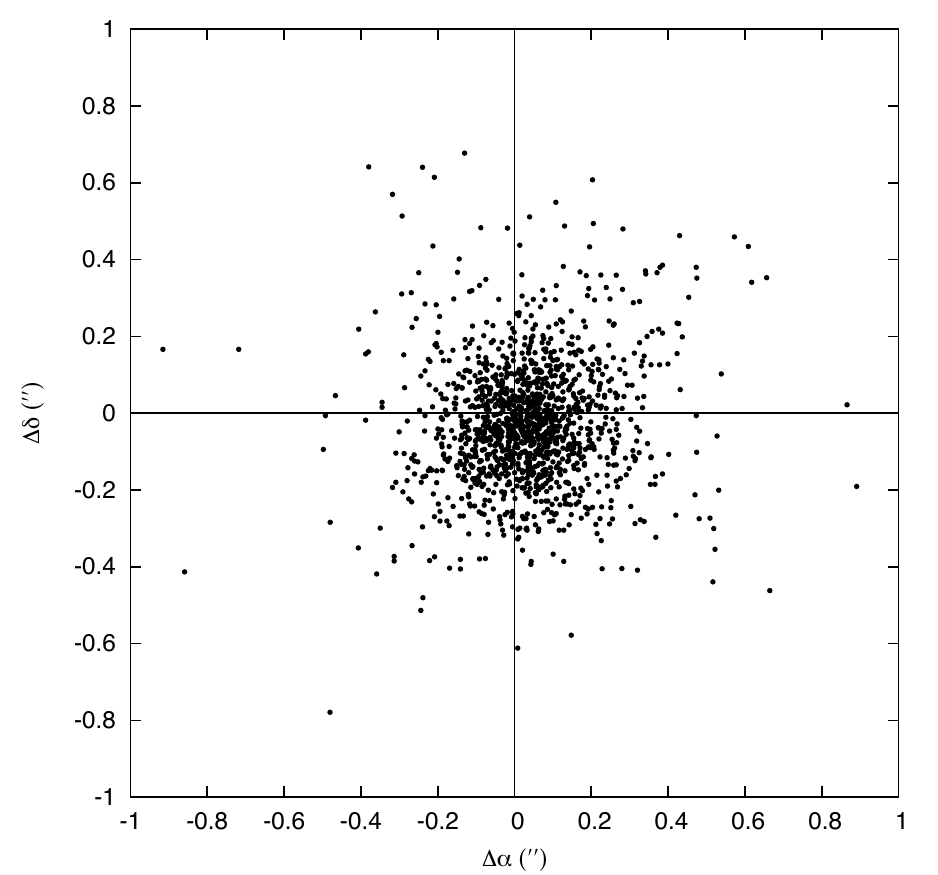}%
 \caption{Plot of $\Delta \delta ''$ versus $\Delta \alpha ''$ for all 
1373 ICRF2 sources in our comparison sample.  Position offsets, $\Delta 
\alpha, \Delta \delta$, are in arcseconds in the sense optical minus 
radio.} 
 \label{ra-dec}
 \end{center}
\end{figure}

\section{RESULTS}\label{results}

{\renewcommand\tabcolsep{3pt}
\begin{table*}[tp]
\caption{Position offsets between SSS and ICRF2, in the sense optical 
minus radio with $\Delta \alpha$ and $\Delta \delta$ measured in arcseconds on the sky.  The defining ICRF2 sources are shown separately. Source 
selection is discussed in section \ref{source_sel}. }
\label{position_offsets}
\begin{center} 
\begin{tabular}{|l|ccccc|ccccc|}
\hline
&&&&&&\\[-0.4cm]
& \multicolumn{5}{c|}{ICRF2 defining sources} &
\multicolumn{5}{c|}{Full ICRF2 sample}\\ 
& $N$ & $\overline{\Delta \alpha}$ $('')$ & $\sigma_\alpha$ & 
$\overline{\Delta \delta}$ $('')$ & $\sigma_\delta$ &
 $N$ & $\overline{\Delta \alpha}$ $('')$& $\sigma_\alpha$
& $\overline{\Delta \delta}$ $('')$ & $\sigma_\delta$\\[0.15cm] 
\hline
&&&&&&\\[-0.4cm]
All sources & 187 & $+0.018 \pm 0.011$ & 0.15 &$-0.013 \pm 0.012$ & 0.17 & 
1373 &
$+0.036 \pm 0.004$ & 0.16 & $-0.025 \pm 0.004$ & 0.16 \\
North $\delta \geq +2.5^{\circ}$ & 107 & $+0.087 \pm 0.019$ & 0.18 & 
$+0.007\pm0.019$ & 0.20 
& 764 &
$+0.089\pm 0.006$ & 0.17 & $-0.012\pm0.006$ & 0.18 \\
South $\delta < +2.5^{\circ}$ & 81 & $-0.026\pm0.015$ & 0.13 & $-0.026 \pm 
0.017$ & 0.16& 
609 &
$-0.031\pm 0.005$ & 0.13 & $-0.041\pm 0.006$ &0.15\\[0.2cm]
\hline
\end{tabular}
\end{center}
\end{table*}   
}

\begin{figure*}[ht]
\begin{center}
\includegraphics[width=\textwidth,clip=]{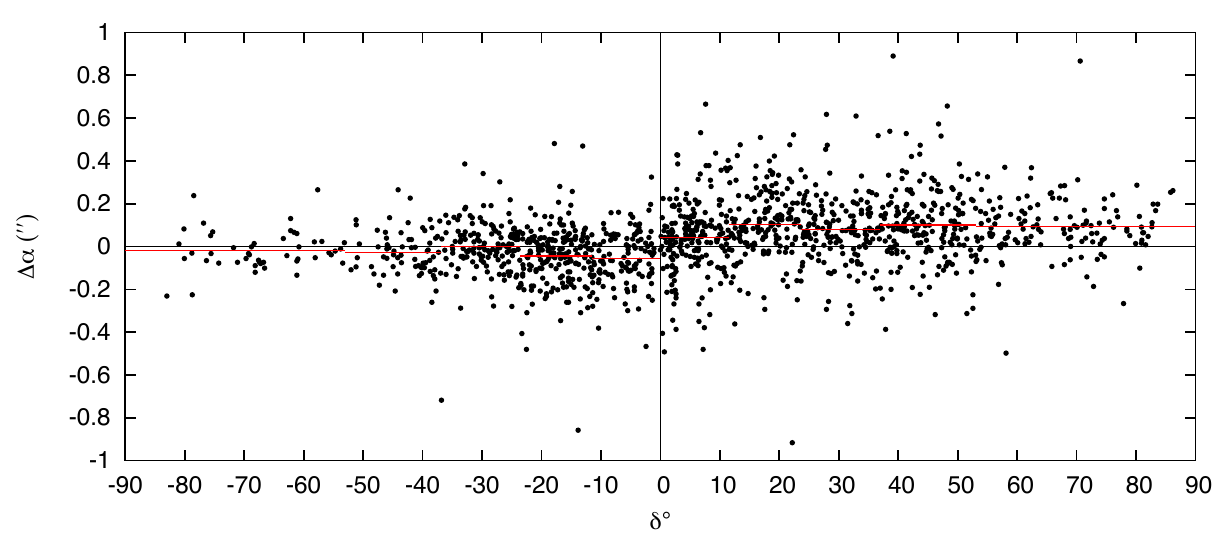}%
\caption{Plot of RA offset (arcsec on the sky, optical minus radio) versus 
declination (deg). The red horizontal bars are group means over intervals 
with comparable numbers of sources.} \label{DRAvsDec}
\end{center}
\end{figure*}

\begin{figure*}[bt]
\begin{center}
\includegraphics[width=\textwidth,clip=]{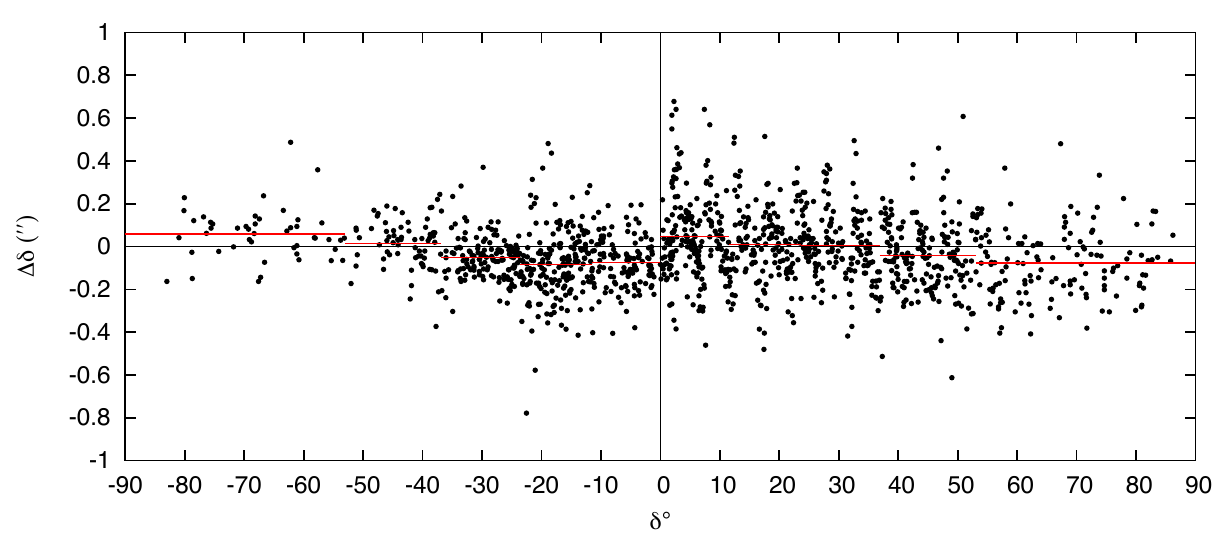}%
\caption{Plot of declination offset (arcsec, optical minus radio) versus 
declination (deg).  The red horizontal bars are group means over 
intervals with comparable numbers of sources.  Note the regular pattern 
of offsets in the north, despite a zero mean offset; this is discussed 
in more detail in section \ref{dec_offsets}.} \label{DDecvsDec}
\end{center}
\end{figure*}

Position differences between ICRF2 and SSS were calculated in arcseconds
on the sky in the sense {\it optical minus radio}. This follows the
convention used by Deutsch \shortcite{deu99} and Hambly et al.\
\shortcite{ham01c}.  The overall distribution of position differences for
the full set of 1373 sources is shown in Fig.\ \ref{ra-dec}.  It is
immediately clear from this cross-plot that there are small but
significant systematic errors in the optical positions, both in RA and
Dec.  On the other hand, given the strong central concentration of points,
it is likely that the small number of sources with radial offsets $>0.5''$
are mostly chance associations.

These results are summarised in Table \ref{position_offsets}, in which
position offsets and errors are given separately for the ICRF2 defining
sources and for the full ICRF2 sample. The two samples are separated
further into northern and southern hemispheres, with the division set at $\delta = +2.5^{\circ}$, the crossover point between the northern and southern surveys.


To investigate these errors further, and separate the contributions from
the northern and southern surveys, we now examine the position
differences as a function of declination in Figs\ \ref{DRAvsDec} and
\ref{DDecvsDec}.  Group means over intervals in declination with
comparable numbers of sources are shown in red.

\subsubsection*{RA offset versus declination, Fig.\ \ref{DRAvsDec}}

The most prominent feature of this plot is the uniform systematic offset
in the north, amounting to $>10 \sigma_{\rm mean}$, as well as a larger
scatter.  The offset is present in all three subsets of ICRF2: defining,
non-defining and VCS.  There is a smaller but significant offset in the
south in the opposite sense.

\subsubsection*{Dec offset versus declination, Fig.\ \ref{DDecvsDec}}

The most striking aspect of this plot is the regular pattern of offsets in 
the north, with peak amplitudes as large as $0.5''$, even though the 
mean offset is not significant.   This highlights the value of our larger 
sample.  No obvious pattern is seen in the south, although there is a 
significant offset of $-0.05''$.

\subsubsection*{RA and Dec offsets versus magnitude}

To investigate the possibility of any systematic effects
occurring as a function of apparent magnitude we have investigated the mean and
scatter in RA and Dec as a function of magnitude. In both the POSS-II and UKST
survey regions there is no significant change in systematic offset at fainter
magnitudes, although the scatter does increase for fainter objects. This is consistent with the results of Hambly et al. \shortcite{ham01c} and Deutsch
\shortcite{deu99}.


\section{DISCUSSION}\label{dec_offsets}

\begin{figure}[h]
 \begin{center}
 \includegraphics[width=\columnwidth,clip=]{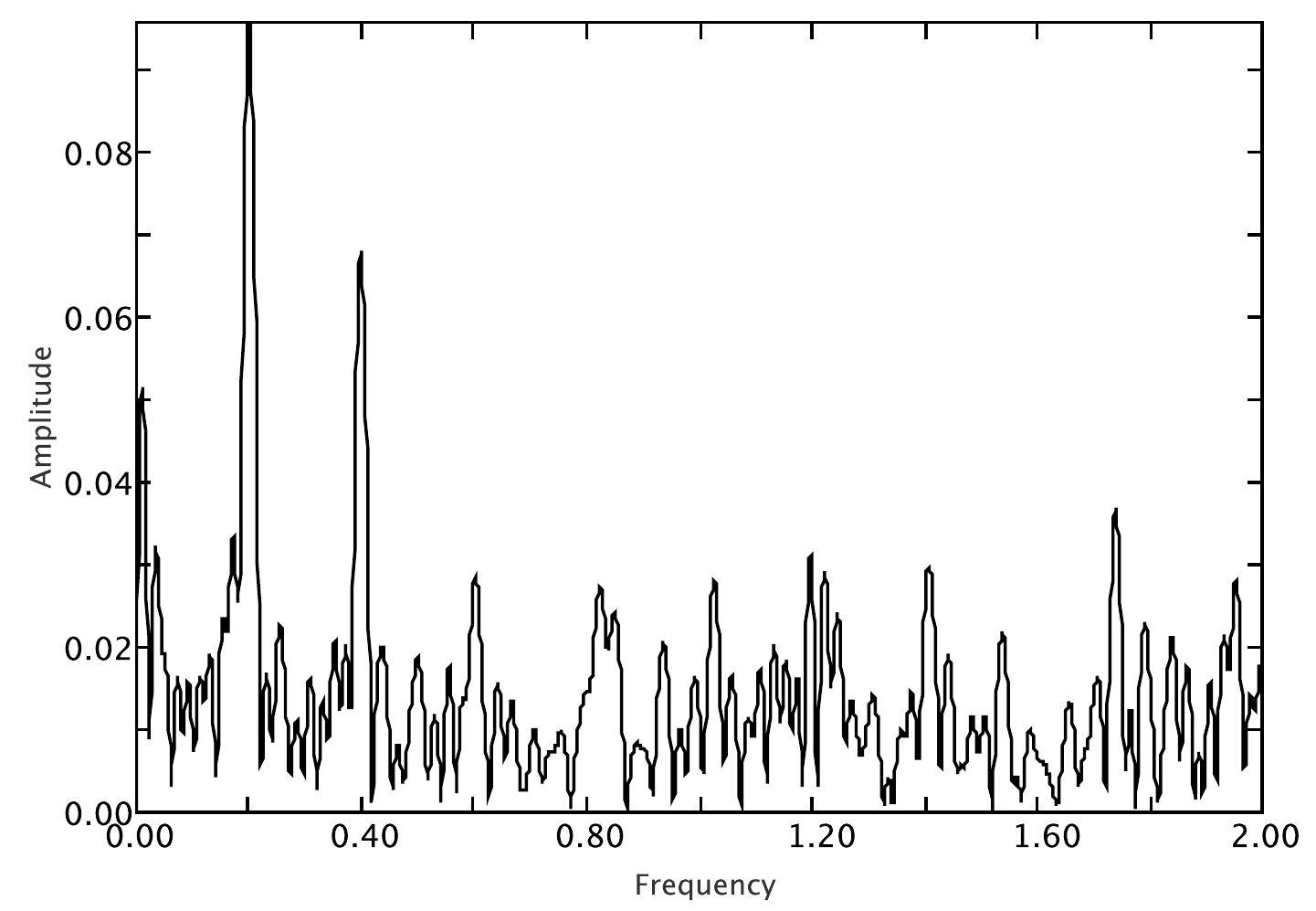}%
 \caption{Power spectrum of the declination offsets between
$+2.5^{\circ}$ and $+52.5^{\circ}$, showing a clear peak at 0.2
deg$^{-1}$, corresponding to the $5^{\circ}$ spacing of plate centres.}
 \label{power_spectrum}
 \end{center}
\end{figure}

Hambly et al.\ \shortcite{ham01c} carried out a detailed set of
comparisons between SSS and ICRF1 in the south, showing that zero-point
errors were no larger than 100 mas for the blue and red surveys.  These
results are confirmed (with substantially larger numbers of sources) by
the results in Table \ref{position_offsets}.  However, a similar analysis
was not done for the northern POSS-II surveys, because they were not
completed in time for inclusion in the definitive papers describing the
SuperCOSMOS Sky Surveys \cite{ham01a,ham01b,ham01c}.  In particular, the
POSS-II J survey scans were only added to the online SSS database in
2008.\footnote{http://www-wfau.roe.ac.uk/sss/status.html}

Extensive checking of the northern declination offsets confirmed that the
repetitive sawtooth pattern was real.  A power spectrum analysis of the
offsets between declinations $+2.5^{\circ}$ and $+52.5^{\circ}$, where the 
effect is most prominent, is shown
in Fig.\ \ref{power_spectrum}.  This reveals a clear peak at a frequency
of 0.20 deg$^{-1}$ or $5^{\circ}$, the spacing of the plate centres,
together with harmonics at 0.4 and 0.6 deg$^{-1}$.  

\begin{figure}[h]
\begin{center}
\includegraphics[width=\columnwidth]{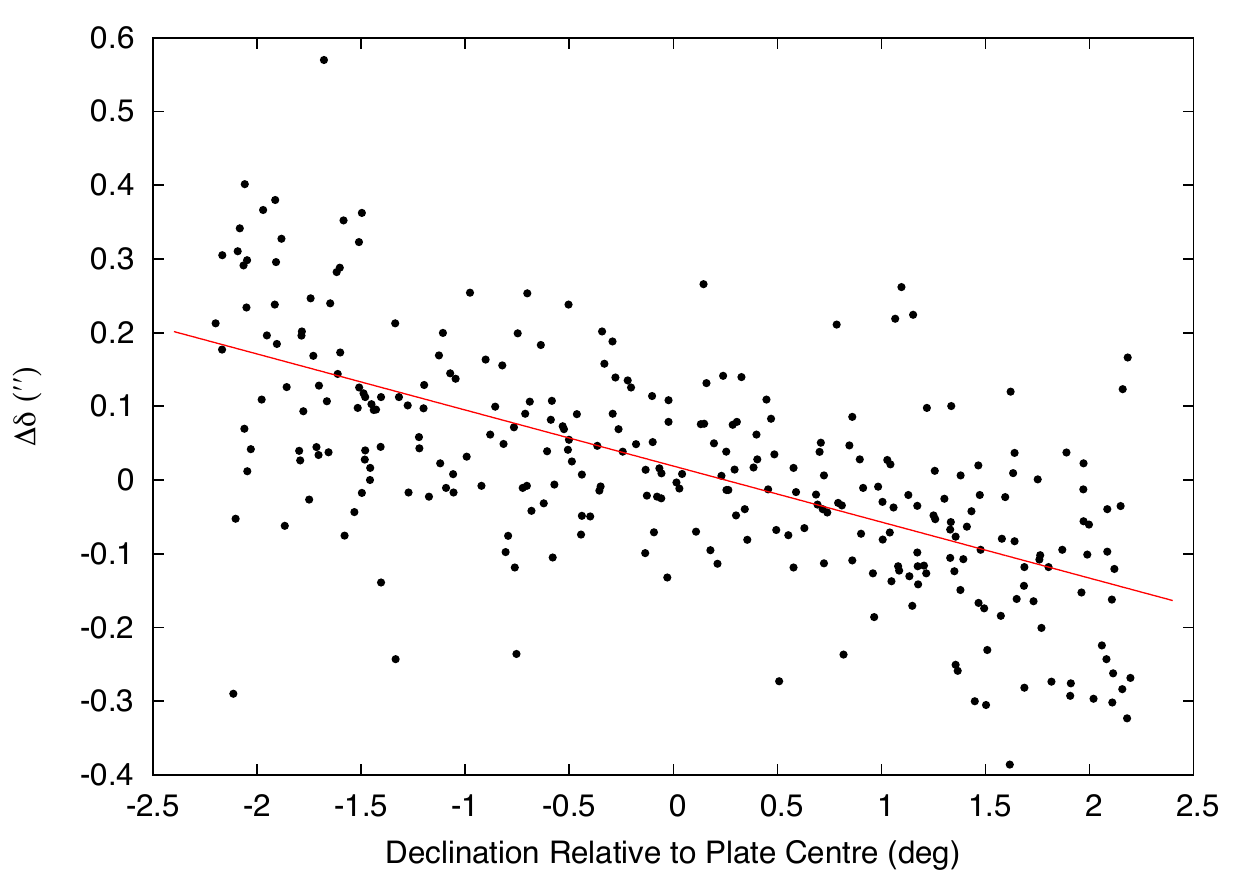}
 \caption{A plot of the declination offset (arcsec, optical
minus radio) for sources with declination in the range $+2.5^\circ < \delta
<+32.5^\circ$, co-added on $5^\circ$ centres. The red line is the least squares
fit to the offsets with slope $-0.076 \pm 0.005 $ and intercept $0.019 \pm
0.006$. }\label{coadd} 
\end{center} 
\end{figure}

The sawtooth pattern was further confirmed by folding and co-adding the offsets
on $5^{\circ}$ centres. Fig.\ \ref{coadd} shows the declination
offsets for sources between declination $+2.5^\circ$ and $+32.5^\circ$ as a
function of distance from the plate centres. Since the plate centres for both
northern and southern surveys were defined in B1950 co-ordinates, while the
optical and ICRF2 positions are defined in J2000, it was necessary to restrict
the plot to $\pm 2.2^{\circ}$ to account for precession in declination between B1950 and
J2000.

Nigel Hambly (2013, personal communication) pointed out that the
mechanical distortions on POSS-II plates \cite{ham01c} were significantly
larger than on the UK Schmidt plates, and suggested that residual
systematics remaining after applying the large ($>1''$)  corrections could
account for the observed pattern in Fig. \ref{DDecvsDec}, which would naturally have a pitch of
$5^{\circ}$.  This argument is supported by the absence of a similar
pattern in the south.
 
 An alternative (or additional) interpretation of Figures\
\ref{DDecvsDec} and\ \ref{coadd} could be differential refraction, possibly
arising from application of the wrong correction for atmospheric refraction
in the original SSS reduction. There is tentative evidence
that the amplitude of the sawtooth pattern increases towards the northern limit
of the POSS-II survey, implying a zenith-angle dependence.  However, it is
difficult to judge whether there is a corresponding dependence north of the
Palomar zenith ($+33^{\circ}$) because of the falling source density and sky
area.

The uniform offset of $0.08''$ in the northern RAs suggests a zero-point
error.  This is surprising since it would have shown up
readily in a comparison with ICRF1, as carried out in Hambly et al.\
\shortcite{ham01c}.  The mean offsets in the south, with fewer sources,
are dominated by the points north of $-20^{\circ}$ and have a smaller
overall scatter than in the north.

\section{CONCLUSION}\label{concl}

A detailed comparison has been carried out between optical positions for
stellar objects from the SuperCOSMOS J surveys carried out with the UK and
Palomar Schmidt telescopes and long-baseline interferometer radio
positions that define the ICRF2 reference frame.  Subject to selection
criteria defined in the paper, we find the overall standard deviations in
the position differences to be $0.16''$ in both coordinates, with
systematic offsets of $<0.1''$.

In the north, we found a uniform systematic offset of $0.09''$ in RA and a
sawtooth pattern with mean amplitude $\sim 0.2''$ in declination, with a pitch of
$5^{\circ}$ corresponding to the spacing of plate centres.  Speculations
on the origin of the periodic offsets in declination include residual
systematics after correction for the mechanical distortions introduced by 
the plateholder, and differential errors in correcting for atmospheric 
refraction.

The overall scatter of the optical positions about ICRF2 is smaller in
the south and systematics, while present, are also smaller.

\begin{acknowledgements}

We thank Oleg Titov for access to the ICRF2 data, Dennis Stello for 
the power spectrum in Fig.\ \ref{power_spectrum}, and Nigel Hambly for discussions on the 
results from this paper.

The Second Palomar Observatory Sky Survey (POSS-II) was made by the
California Institute of Technology with funds from the National Science
Foundation, the National Aeronautics and Space Administration, the
National Geographic Society, the Sloan Foundation, the Samuel Oschin
Foundation, and the Eastman Kodak Corporation. The Oschin Schmidt
Telescope is operated by the California Institute of Technology and
Palomar Observatory. The UK Schmidt Telescope was operated by the Royal
Observatory Edinburgh, with funding from the UK Science and Engineering
Research Council (later the UK Particle Physics and Astronomy Research
Council), until 1988 June, and thereafter by the Anglo-Australian
Observatory. The blue plates of the southern Sky Atlas and its Equatorial
Extension (together known as the SERC--J/EJ) were taken with the UK
Schmidt Telescope.  All data retrieved from
URLs described herein are subject to the copyright given in this
copyright summary. Copyright information specific to individual
plates is provided in the downloaded FITS headers.

\end{acknowledgements}


\end{document}